# Exciton self-trapping causes picoseconds recombination in metal-organic chalcogenides hybrid quantum wells

Christoph Kastl[†§*], Adam M. Schwartzberg[†*], and Lorenzo Maserati[†‡*]

† The Molecular Foundry, Lawrence Berkeley National Laboratory, Berkeley, California 94720, USA.

§ Walter Schottky Institute and Physik Department, Technical University of Munich, Garching 85748, Germany.

‡ Center for Nano Science and Technology @PoliMi, Istituto Italiano di Tecnologia, 20133 Milan, Italy.

AUTHOR INFORMATION

Corresponding Authors

* christoph.kastl@wsi.tum.de, amschwartzberg@lbl.gov, lmaserati@lbl.gov.

ABSTRACT

Metal-organic species can be designed to self-assemble in large-scale, atomically defined, supramolecular architectures. Hybrid quantum wells, where inorganic two-dimensional (2D) planes are separated by organic ligands, are a particular example. The ligands effectively provide an intralayer confinement for charge carriers resulting in a 2D electronic structure, even in multilayered assemblies. Air-stable metal organic chalcogenides hybrid quantum wells have recently been found to host tightly bound 2D excitons with strong optical anisotropy in a bulk matrix. Here, we investigate the excited carrier dynamics in the prototypical metal organic chalcogenide $[AgSePh]_\infty$, disentangling three excitonic resonances by low temperature transient absorption spectroscopy. Our analysis suggests a complex relaxation cascade comprising ultrafast screening and renormalization, inter-exciton relaxation, and self-trapping of excitons within few picoseconds. The ps-decay provided by the self-trapping mechanism may be leveraged to unlock the material's potential for ultrafast optoelectronic applications.

**Introduction**

Atomically thin, inorganic semiconductors have gained enormous research interest as a test bench for room temperature manipulation of two-dimensional (2D) Wannier excitons.[1] Semiconductor hybrid quantum wells, such as metal-organic chalcogenides (MOCs) or metal halide perovskites,[2,3] show similar 2D excitonic resonances with binding energies up to hundreds of meV. In these hybrid materials, the inorganic planes are spaced by organic ligands, which provide a low dielectric screening environment and quantum confinement for the charge carriers. Due to the low dielectric screening, the optical properties are dominated by strongly bound exciton complexes.[4] This marks a stark contrast to all-inorganic quantum wells (e.g. III-V semiconductors-based), where both the insulating barriers and the active layers have high dielectric constants and diminish the exciton binding to few meV, restricting studies and applications of excitonic phenomena to cryogenic temperatures.[5] Hybrid, multilayered structures open an avenue for combining the chemical tunability of organic chemistry, bulk material systems suitable for scalable thin film technologies, and the advantageous (opto-)electronic properties of low-dimensional inorganic materials. This innovative concept of hybrid materials with strong excitonic character is rapidly emerging from the converging fields of metal-organic frameworks,[6,7] metal-organic coordination polymers,[8] and reduced dimensionality perovskites.[2] While two-dimensional metal halide perovskites are regarded as particularly appealing and successful proof-of-concept, their applicability remains limited, because their ionic lattice is critically affected under standard operating conditions.[3] Air-stable MOCs with covalent character constitute therefore an attractive alternative for providing a hybrid quantum well platform for next generation optoelectronics.[4,9] So far, tens of different MOCs featuring quantum wire and quantum well nanostructures,[10–18] potentially hosting low-dimensional physics phenomena, were reported. However, research on this material platform has focused mostly on crystallographic and structural aspects, while their optoelectronic properties have been largely unexplored, with notable exceptions.[13–18]

An atomistic representation of [AgSePh]$_\infty$ (silver benzeneselenolate), a prototypical member of the family of layered MOCs, is shown in Fig. 1a. Its structure appears as a stack of inorganic/organic layers, with a periodicity of ~14 Å along the out-of-plane direction. Similar to layered transition metal chalcogenides, the individual layers are only weakly bound in the out-of-plane direction (*c*-axis) by van der Waals-type interactions between the benzene functional groups. Although the crystal structure was known for almost two decades,[19] its optical and electronic properties were only recently elucidated.[4,9,20] A combined theoretical and experimental study showed that [AgSePh]$_\infty$ is a direct gap semiconductor with large optical anisotropy both between the out-of-plane *c*-axis and the in-plane *ab*-axes, as well as strong anisotropy within the *ab*-plane.[4] The optical absorption is dominated by tightly bound intralayer excitons, which are confined within the 2D silver-selenide planes. At room temperature, the main excitonic transitions occur at 2.672 eV (464 nm, $X_1$), 2.738 eV (453 nm, $X_2$) and 2.870 eV (432 nm, $X_3$), and the single particle gap was determined to be 3.05 eV (407 nm) corresponding to an exciton binding energy of 380 meV, for the lowest excitation $X_1$.[4,20] A recent study investigated the excitonic fine structure via linear and non-linear photoluminescence excitation spectroscopy, adding several dark states as well as a sub-bandgap luminescence at low-temperatures to the picture.[20] From the time-resolved luminescence decay, an excited state lifetime on the order of 20 ps was deduced, although the origin of the fast luminescence decay was not determined.[20] Such ps-lifetimes can be appealing for high bandwidth optoelectronic applications, but this demands also precise knowledge of the fundamental decay mechanism and the involved energy levels. Here, we use ultrafast, broadband transient absorption spectroscopy to unravel the excited state dynamics in [AgSePh]$_\infty$ from femtoseconds to nanoseconds. We find a complex relaxation cascade involving screening and band-gap renormalization, inter-excitonic relaxation, trapping of excitons, and non-radiative coupling to the phonon bath. As the main finding of our analysis, we show that the dynamics are governed by fast self-trapping of charge carriers and/or excitons with timescales below 10 ps. Carrier trapping processes are of central importance for the utilization of excitonic materials in optoelectronic applications. While for energy conversion processes trapping is usually considered detrimental,[21] it can open new pathways for utilizing low-dimensional materials, and in particular [AgSePh]$_\infty$, for example as ultrafast optical switches, high-bandwidth detectors, or broadband pulsed light sources.[22,23]

**Results**

Thin films were prepared on quartz substrates by a chemical vapour–solid reaction reported previously yielding nanocrystalline films with the *ab*-plane oriented mostly parallel to the substrate.[4,24] Due to the nanocrystalline morphology (grain size ~200 nm), the in-plane anisotropy was not relevant for all following experiments (spot sizes ~ 150 μm). Figure 1b displays steady state absorbance spectra of a 70 nm thin film from room temperature to 77 K. At low temperatures, the absorbance peaks significantly narrow, and the three excitonic resonances $X_1$, $X_2$, and $X_3$ can clearly be discerned. Additionally, an onset feature develops at 3.13 eV (396 nm), which is close to the energy of the single-particle gap determined from previous experiments at room temperature.[4,20] The transient absorption spectrum (Fig. 1c) also reveals these three distinct resonances directly after optical excitation. The measurements were done in a transmission geometry, and we report the transient change in transmission -$\Delta T/T$. The latter corresponds approximately to the change in absorption, when neglecting interference and reflection effects for very thin films on transparent substrates.[25] Figure 1d shows the broadband transient absorption dynamics at a bath temperature of 77 K for an excitation wavelength of 3.758 eV (330 nm). As will be discussed in the following, the main transient changes of the excitonic resonances can be understood by ultrafast ground state depletion as well as bandgap renormalization due to the excited carrier ensemble on very short time scales ($t < 0.5$ ps), and by a heated phonon bath on long time scales ($t \sim 100$ ps). The decay of the excitonic features on intermediate time scales ($t < 10$ ps) is accompanied by the emergence of a broad, sub-gap photoinduced absorption (PIA). The latter is emphasized in Fig. 1d by the adjusted colour scale for wavelengths larger than 480 nm. While in steady-state absorption such a sub-gap feature is absent, low-temperature luminescence clearly shows a corresponding broadband emission centred near 600 nm (Supplementary Fig. S1). Since the decay time of the excited excitonic features matches closely the rise time of the PIA (Fig. 1e), we assign the PIA to absorption from a trapped state populated by the decay of the exciton ensemble. Trapping can be relevant both for free carriers, i.e. the electron and/or hole are trapped independently, or excitons, i.e. the bound electron-hole pair is trapped directly.[21] The microscopic trapping mechanism is often mediated by defects, which serve as localization sites for the carriers (extrinsic trapping).[26] For soft lattice configurations, as is the case for layered [AgSePh]$_\infty$, so-called intrinsic self-trapping can often become particularly relevant.[27] Then, the localization of the charge carrier or exciton at the trapping site induces a lattice deformation or even, in some cases, a reconfiguration of the local bonds,[28–30] providing a deep potential well for the localized state.

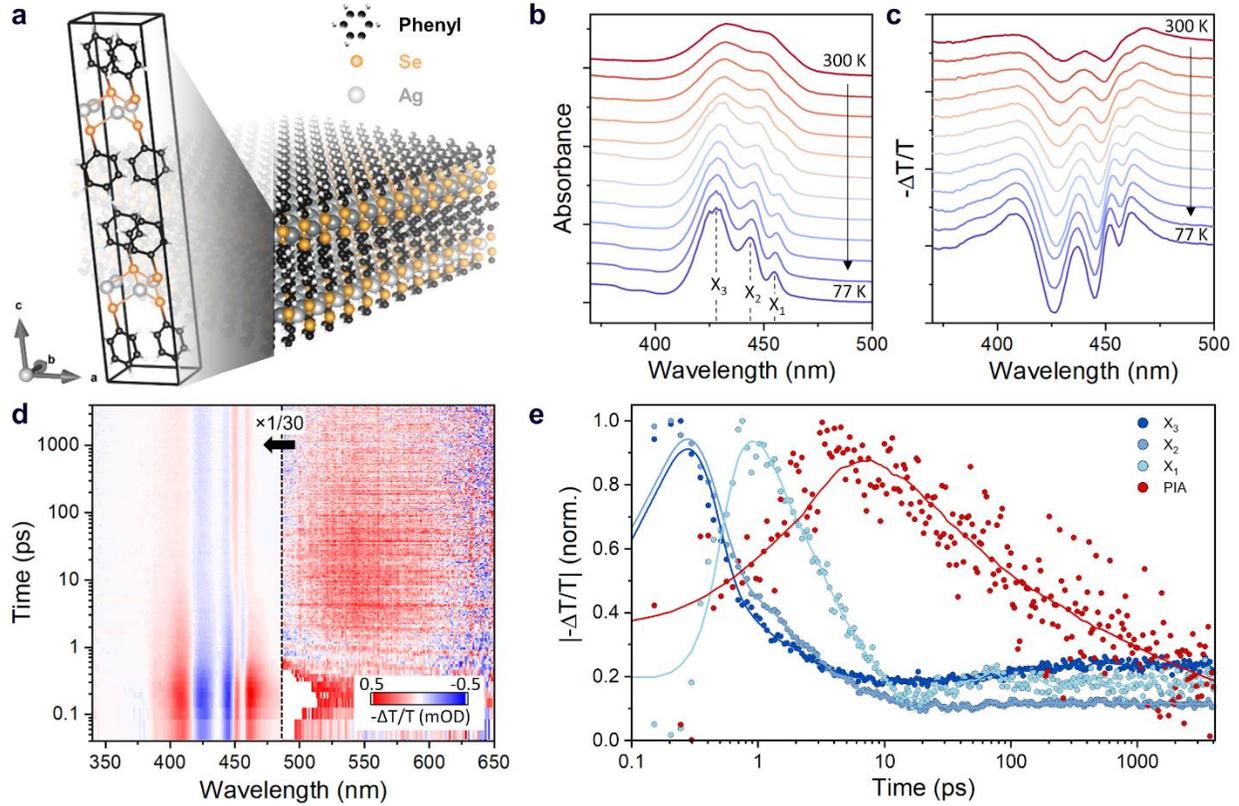

**Figure 1a,** Atomistic representation of [AgSePh]$_\infty$. **b,** Thin film absorption spectra and **c,** transient absorption spectra directly after pump excitation from room temperature down to 77 K. The three excitonic resonances are labelled $X_1$, $X_2$ and $X_3$. **d,** Broadband transient absorption map recorded in transmission geometry (-$\Delta T/T$). Below the optical gap, the colour scale was modified as indicated to highlight the spectrally broad photoinduced absorption (PIA) centred around 550 nm (pump wavelength 330 nm, pump energy 0.7 µJ, $T_{bath}$ = 77 K). **e,** Dynamics extracted at 425 nm ($X_3$), 445 nm ($X_2$), 456nm ($X_1$), and 550 nm (PIA). The signal of $X_1$ (light blue dots) is delayed by about 500 fs with respect to $X_2$ and $X_3$ (blue and dark blue dots). The rise of the PIA (red) matches closely the decay of the excitonic features. The solid lines are smoothed to highlight the trend.

To better disentangle the effect of free carriers vs. excitons in the trapping process, we turn to resonant excitation of the excitons (Figure 2, pump wavelength 425 nm, $T_{bath}$ = 6 K). We distinguish three distinct temporal regimes (labelled I,II, and III). For $t$ < 600 fs (regime I), the dynamics are governed by an initial, ultrafast relaxation of the non-equilibrium carrier population (Fig. 2b). While the photoinduced bleaching signals (minima in -$\Delta T/T$) of $X_3$ and $X_2$ decay, the photoinduced bleaching of $X_1$ rises (highlighted by the arrow in Fig. 2b), indicating relaxation and energy transfer of the excited carriers into the lowest transition $X_1$. The latter is also apparent as a distinct kink of $X_1$ in the transient absorption spectral map (highlighted by the arrow in Fig. 2a).

We resolve a further transient feature at 385 nm, which matches closely the step-like onset found in the steady-state absorption (cf. Fig. 1b). The latter may correspond either to the onset of the continuum (significantly blue-shifted compared to room temperature) or a higher excitonic transition.

In addition to photoinduced bleaching, the transient response comprises also spectral shifts of the exciton transitions, clearly visible by the combined positive and negative transmission changes. For 2D excitons in layered materials, screening and band gap renormalization effects can play a dominant role for the ultrafast excitonic response,[31,32] and they have been shown to lead to spectral shifts of the optical transitions. In control experiments, where we tuned the excitation wavelength (Supplementary Fig. S2), we found that also the higher lying transitions ($X_3$ and $X_2$) shift when resonantly pumping only the lowest exciton transition $X_1$. Such a collective shift of all transitions, independent of the excitation wavelength, can be understood by bandgap renormalization and screening due to the photogenerated charge carriers, as has been shown for the case of $MoS_2$, $WSe_2$, and other 2D materials.[31,33–35] This hypothesis is further corroborated by the fact that the relaxation times extracted at the position of $X_1$, $X_2$ and $X_3$ are similar for all pump wavelengths (Supplementary Fig. S3), suggesting a common origin. Generally, the interplay between the renormalizations of single particle gap and exciton binding energy can be complex.[1] For the case of 2D transition metal dichalcogenides, it has been shown that they compensate to a large extent such that the overall shifts of the optical transitions are rather small (on the order of 10 meV) while the change in single particle gap (or equivalently exciton binding energy) may be rather large (on the order of 100 meV).[36]

From 0.6 ps < $t$ < 10 ps (regime II, Fig. 2c), the transitions associated with $X_1$, $X_2$, and $X_3$ decay with similar time constants, and the photoinduced absorption (PIA centred at 550 nm) arises, similar to the case of free carrier pumping (cf. Fig. 1d and Supplementary Fig. S4). Below, we provide further evidence that these dynamics are governed by direct self-trapping of excitons. For $t$ > 10 ps (regime III, Fig. 2d), trapping is essentially complete and the transient absorption signatures can be described solely by temperature-induced changes of the steady state absorbance spectrum d$A$/d$T_{bath}$. In other words, any additional, non-radiative relaxation pathways have led to a heated phonon bath, which in turn affects the excitonic absorption (cf. Fig. 1b). By comparing $\Delta T/T$ and d$A$/d$T_{bath}$, we can estimate an increase of the lattice temperature by 0.9 K at $t$ = 4 ns for a 0.04 µJ pulse (Supplementary Fig. S5). The phonon temperature exhibits a nanosecond decay past our instrumental limit, consistent with cooling to the quartz substrate. For samples supported on silver films, where heat conduction is more efficient, we find a much faster decay (340 ± 60 ps). Importantly, although the PIA band still persists for $t$ > 10 ps, it is completely absent in the d$A$/d$T_{bath}$ spectrum, and we exclude temperature effects to be its origin.

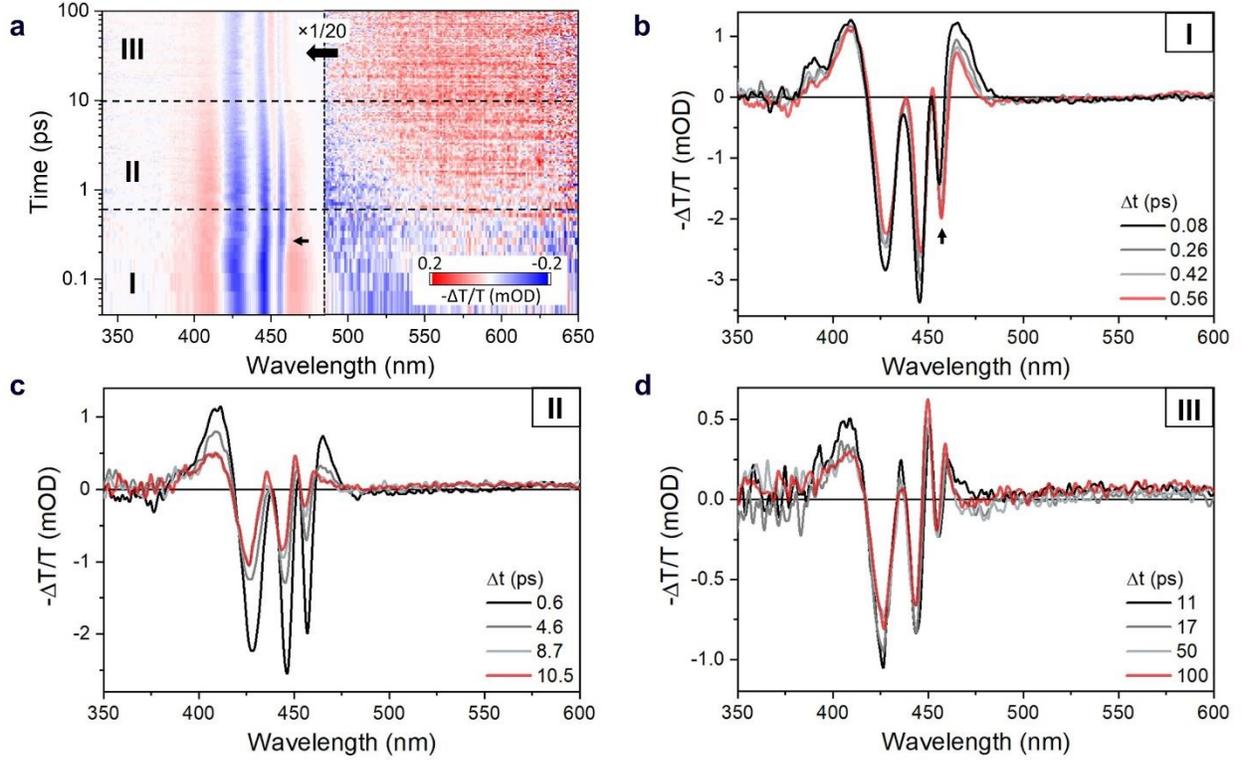

**Figure 2.** Transient absorption spectroscopy for excitonic pumping at 6 K. **a,** Broadband TA map in logarithmic representation. Three distinct temporal regimes can be discerned as indicated by dashed lines corresponding to thermalization and relaxation (I), exciton self-trapping (II), and phonon cooling to the substrate (III). For wavelengths above 480 nm, the colour scale was adjusted as indicated. **b-d,** Transient absorption spectra for specific time delays in the different regimes (pump wavelength 425 nm, pump energy 0.015 µJ, $T_{bath}$ = 6 K).

To further elaborate the decay dynamics, we depict the pump fluence dependence in Fig. 3, both for resonant excitonic pumping of $X_3$ (pump wavelength 429 nm, Fig. 3a-c) and free carrier pumping (pump wavelength 330 nm, Fig. 3d-f). Figure 3a shows the spectral response at $t$ = 150 fs normalized by the pump fluence. When pumping below the quasiparticle gap, the spectra comprise clear transient bleaching signatures of all three excitons (negative peaks in Fig. 3a). Figure 3b shows the pump fluence dependence evaluated at the position of the negative peaks. As expected, the transition $X_3$, which is the one being pumped directly, shows an almost linear behaviour (slope 0.94 ± 0.01), and it saturates only for the highest pump fluences > 0.2 µJ, equivalent to an average exciton density per layer > $3 \cdot 10^{13}$ cm$^{-2}$. We estimated the latter from the measured steady-state absorbance spectrum and the film thickness. We note that for the largest pump fluences, alongside possible Auger-assisted recombination,[37,38] slight degradation of the films occurred, which may also cause the sublinear trend.

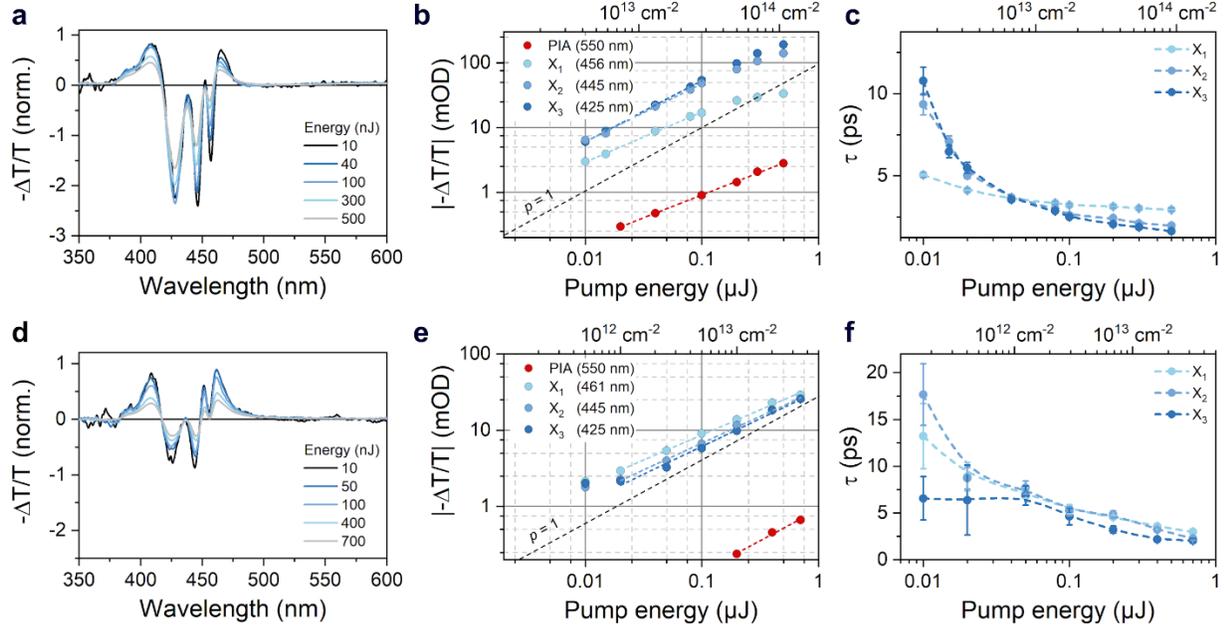

**Figure 3.** Pulse fluence dependence of the TA for above bandgap (pump wavelength 330 nm) **(a-c)** and excitonic pumping (pump wavelength 429 nm) **(d-e)** at $T_{bath}$ = 77 K. **a,d,** Instantaneous TA signal after the pump normalized by pump fluence. **b,e,** Maximum TA signal vs. pump fluence for the three excitonic features peaks and the photoinduced absorption band (PIA) in double logarithmic representation. The dashed lines are power law fits. The black dashed line indicates the slope of a power law with exponent p = 1. **c,f,** Time constants extracted from an exponential fit vs. pump fluence for $X_1$, $X_2$, $X_3$ and the PIA band. The top axis labels represent the effective exciton density per layer estimated from the steady state absorbance and film thickness.

The lowest transition $X_1$ is markedly different with a clear sublinear dependence (0.77 ± 0.01) starting already for the lowest pump energies of 0.01 µJ (corresponding to an average exciton density per layer of 6·$10^{11}$ cm$^{-2}$). The PIA at 550 nm exhibits a similar sublinear dependence (slope 0.71 ± 0.02) as well. The common slopes suggest that the 550 nm band is populated from $X_1$, in other words, that $X_1$ is trapped. The sublinearity of $X_1$ furthermore suggests that in addition to the fast population of $X_1$ via $X_3$ (~500 fs), another competing process can relax the higher excitonic states, which is likely direct trapping of $X_2$ and $X_3$. We note that a sublinear dependence of the trapped exciton signature is commonly attributed to extrinsic self-trapping, i.e. defect-mediated trapping.[21,39] In this picture, the sublinearity is caused by the saturation of the finite defect density at high fluences. However, in the present case, $X_1$ shows already a sublinear dependence, such that we cannot unambiguously assign the sublinear slope of PIA to a defect mediated process. Furthermore, we find similar signatures of self-trapping in single crystal samples (Supplementary Fig. S6), which we expect to have considerably lower defect density than the nanocrystalline films. Therefore, trapping of excitons is likely an intrinsic property of [AgSePh]∞.

In Fig. 3c, we plot the decay times extracted from an exponential fit. We note that the assumption of exponential dynamics is not strictly correct for multiple level systems, since the coupled rate equations can generally result in non-exponential behavior.[40] However, we find that the decay dynamics can be satisfactorily approximated by an exponential model (Supplementary Fig. S7 and Supplementary Fig. S8). The relaxation time of $X_1$ is 3 ps < $\tau_{X1}$ < 5 ps and it depends only weakly on excitation density. An independent study estimated the radiative relaxation rate of $X_1$ in [AgSePh]$_\infty$ single crystals to be on the order of 20 ps at low temperature, consistent with a radiative quantum yield on the order of few percent.[20] Therefore, $\tau_{X1}$ represents predominantly the non-radiative decay rate via trapping. The fluence dependence of $X_2$ and $X_3$ suggests that $\tau_{X2}$, $\tau_{X3}$ are impacted by a carrier density dependent mechanism. At large 2D exciton densities, exciton-exciton interactions may become relevant. For example, in the context of 2D transition metal dichalcogenides, experiments evidenced an excitonic Mott transition around $10^{13}$ cm$^{-2}$,[41,42] and theoretical work even suggests that depending on the residual carrier density as well as the dielectric screening a balance between exciton fusion and fission can lead to coexistence of free carriers and excitons already at densities substantially below the Mott transition.[43] Such presence of free carriers may explain features related to spectral shifts, such as the photoinduced absorption (positive peak) at 410 nm in Fig. 3a. Furthermore, in the presence of free carriers, Auger recombination is known to be an effective relaxation pathway, both for the excitons as well as the free carriers, and it may contribute to the observed density dependence of the dynamics of $X_2$ and $X_3$.[37,38]

Figures 3d-f depict the case of free carrier pumping above the quasiparticle gap (pump wavelength 330 nm). Notably, the spectra in Fig. 3a comprise mainly shift signals, i.e. the spectral response integrated across the oscillators averages to zero. These shift signals can be understood by a dominating effect of band gap renormalization due to screening from the free carriers. In a simplistic model, it is expected that the screening due to free carriers is twice as large compared to bound excitons.[33,44] Such screening effects should impact all transitions alike, which is consistent with the common power law (Fig. 3e) as well as common decay time (Fig. 3f) for $X_1$, $X_2$ and $X_3$ within the experimental uncertainty.[31] Importantly, the maximum intensity of the PIA is reduced approximately by a factor of 2 (red dots in Fig. 3b-e and Supplementary Fig. S9) compared to resonant pumping (at the equal density of optically generated electron-hole pairs). The latter decrease corroborates that direct trapping of excitons rather than trapping of free carriers is the dominant mechanism. The overall timescales and spectral responses of the exciton system deduced above are independently corroborated by a global fit analysis of the transient absorption data (Supplementary Fig. S10).

Finally, we discuss the temperature dependence of the carrier dynamics. Overall, we find that the exciton dynamics are only weakly temperature dependent, but show a systematic non-monotonous behaviour (Fig. 4a and Supplementary Fig. S11). The dynamics accelerate slightly as the temperature is decreased, and for temperatures below 50 K, the exciton lifetime appears to increase again. Importantly, the rise time of the STE, as extracted from the maximum of the PIA, closely mimics the temperature dependence of the exciton decay, underscoring the connection between the exciton decay and filling of the trapped exciton states. The decrease of exciton lifetime with decreasing temperature suggests that phonon scattering is not the limiting factor for exciton relaxation. Phonon scattering would generally show the opposite trend, with longer lifetimes at lower temperatures due to a reduction of phonon-assisted non-radiative recombination.[45] By contrast, a combination of exciton self-trapping and dynamic detrapping can explain the observed, non-monotonic temperature dependence.

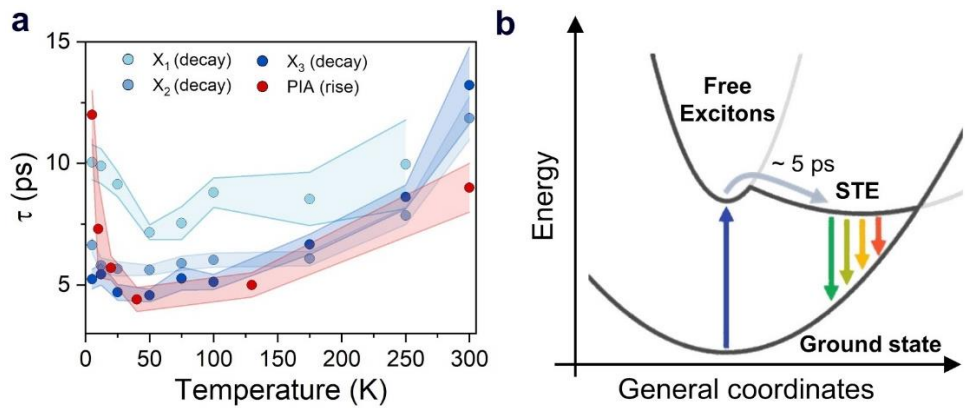

**Figure 4. a,** Temperature dependence of exciton dynamics. The exciton decays are weakly temperature dependent, but follow closely the trapping dynamics. Below 50 K, the thermal energy is not sufficient for promoting the exciton trapping process above the activation energy, causing an increase of exciton lifetime. The shaded areas denote the uncertainty of the parameters (pump wavelength 330 nm, pump energy 0.02 µJ). **b,** Schematic general coordinate diagram of the excitonic transitions, the trapping barrier, and the final self-trapped exciton state. The blue arrow denotes the optical excitation of free excitons. The coloured arrows indicate possible radiative recombination pathways of the trapped state.

Without considering the details of the underlying microscopic process, the trapping process of free excitons can be regarded in a general coordinate representation (Fig. 4b). To reach the self-trapped state an energy barrier must be overcome, e.g. a deformation of the lattice or the bonds at the localization site.[46] The corresponding energy barrier for the inverse detrapping process is typically larger, because the final trapped state, after intermediate relaxation, has an overall lower energy configuration (Fig. 4b).[46] With decreasing bath temperature, the detrapping probability decreases faster than the trapping probability, leading to an overall higher trapping rate. In our interpretation, when the temperature falls below 50 K, the trapping process becomes the slowest, rate-limiting step, and, in turn, the exciton lifetime increases again.[46] A similar, non-monotonic behaviour was observed in photoluminescence spectroscopy of [AgSePh]$_\infty$, yet the impact of self-trapping was not elucidated.[20] The trapped exciton can then further relax to its minimum energy configuration, which is described by general coordinates displaced from the free exciton configuration. As depicted in Fig. 4b, radiative recombination from the self-trapped excited state to the ground state can occur via a continuum of allowed transitions, consistent with the broad sub-gap photoluminescence observed at low-temperatures (Supplementary Figure S1).

**Conclusions**

For the hybrid quantum well material [AgSePh]$_\infty$, we found spectral signatures of three sharp, excitonic transitions $X_1$, $X_2$ and $X_3$, which are also present in the steady state absorption spectrum, as well as a broad photoinduced absorption below the optical gap, which is only present after optical excitation. The transient excitonic features can be assigned to a combination of bandgap renormalization and ground state depletion induced by the excitons and free carriers generated upon excitation. For pumping of free carriers above the optical gap, we determined the formation time of the lowest lying exciton $X_1$ to be on the order of 500 fs. As a main finding, we identified the sub-gap, photoinduced absorption as the signature of self-trapped excitons in this material. The self-trapped state can be more efficiently populated by direct pumping of excitons compared to pumping of free carriers, indicating direct, intrinsic excitonic trapping as the underlying microscopic mechanism. At cryogenic temperatures, the trapped excitons are very likely the origin of broadband sub-gap emission, observed also previously.[4,20] For temperatures below 50 K, we found a decrease of the self-trapped exciton population, which we assign to a lower probability of thermal activation across the energy barrier associated with the trapping process. Overall, the trapping is fast and depopulates the excited carriers within a few picoseconds. We anticipate that [AgSePh]$_\infty$ may find applications as an optical ultrafast switch, and that the self-trapping could be used in general as a driving concept for achieving ultrafast, charge neutral de-excitation in similar material systems.

**SUPPORTING INFORMATION**

The SI are available online.

**METHODS**

**Synthesis of [AgSePh]$_\infty$ (silver benzeneselenolate).** The materials were synthesized according to a protocol previously described in Ref. 4. Briefly, 10 nm thin films of silver were thermally evaporated on fused silica substrates. *1.* To obtain nanocrystalline films, the Ag films were exposed to UV ozone for 10 min to oxidize silver. The oxidized silver films were reacted with benzeneselenol (PhSeH) by vapour transport for 2 hours in an oven at 80 °C. *2.* To obtain microns-size crystals for single crystal experiments, the Ag films were directly reacted with PhSeH for 4 days in an oven at 80 °C. All the samples were rinsed in IPA to remove the unreacted benzeneselenol and $N_2$ dried.

**Steady state absorption spectroscopy.** Temperature dependent absorption spectra were obtained by measuring transmission spectra on a commercial PerkinElmer Lambda 1050 UV/Vis/NIR spectrometer. The samples were mounted in a nitrogen flow cryostat (Linkam THMS350V).

**Photoluminescence spectroscopy.** Low temperature photoluminescence was collected by a confocal microscope (inVia Raman Microscope Renishaw) using a 20X objective and an excitation wavelength of 442 nm. The samples were mounted in a nitrogen flow cryostat (Linkam THMS350V).

**Transient absorption spectroscopy.** We used an ultrafast transient absorption system with a tuneable pump and white-light probe to measure the differential transmission through thin silver benzeneselenolate films supported on quartz substrates. The laser system consists of a regeneratively amplified Ti:sapphire oscillator (Coherent Libra), which delivers 4 mJ pulse energies centred at 800 nm with a 1 kHz repetition rate. The pulse duration of the amplified pulse is approximately 80 fs. The laser output was split by an optical wedge to produce the pump and probe beams and the pump beam wavelength was tuned by an optical parametric amplifier (Coherent OPerA). The pump beam was focused onto the sample by spherical lens at near-normal incidence (spot size FWHM ~ 300 µm). The probe beam was focused onto a sapphire plate to generate a white-light continuum probe, which was collected and refocused onto the sample by a spherical mirror (spot size FWHM ~ 150 µm). The transmitted white light was collected and analysed with a commercial absorption spectrometer (Helios, Ultrafast Systems LLC). Pulse-to-pulse fluctuations of the white light continuum were accounted for by a simultaneous reference measurement of the continuum. Pump and probe beam were linearly cross-polarised and any scattered pump-light into the detection path was filtered by a linear polarizer. The time delay was adjusted by delaying the pump pulse with a linear translation stage (minimum step size 16 fs). The sample was mounted directly on the cold finger of a He-flow cryostat (Janis ST-500) by silver paste. The bath temperature was measured at the cold finger. All measurements were conducted in vacuum ($p \sim 10^{-5}$ mbar).

## AUTHOR INFORMATION

The authors declare no competing financial interests.

## ACKNOWLEDGMENT

Work at the Molecular Foundry was supported by the Center for Novel Pathways to Quantum Coherence in Materials, Office of Science, Office of Basic Energy Sciences, of the U.S. Department of Energy under contract no. DE-AC02-05CH11231. L.M. thanks D. Cortecchia for helping with low-temperature steady state absorption measurements.

## AUTHOR CONTRIBUTIONS

C.K., A.M.S., and L.M. conceived and designed the experiments. L.M. synthesized the material. C.K. and L.M. performed transient absorption spectroscopy at the Molecular Foundry. L.M. performed optical spectroscopy at IIT. C.K. and L.M. analysed the experimental data and wrote the manuscript with input from all authors. All authors reviewed the manuscript.

## REFERENCES

1.  Wang, G. *et al.* Colloquium: Excitons in atomically thin transition metal dichalcogenides. *Rev. Mod. Phys.* **90**, 021001 (2018).
2.  Chen, Y. *et al.* 2D Ruddlesden–Popper Perovskites for Optoelectronics. *Adv. Mater.* **30**, 1703487 (2018).
3.  Blancon, J.-C., Even, J., Stoumpos, Costas. C., Kanatzidis, Mercouri. G. & Mohite, A. D. Semiconductor physics of organic–inorganic 2D halide perovskites. *Nat. Nanotechnol.* **15**, 969–985 (2020).
4.  Maserati, L. *et al.* Anisotropic 2D excitons unveiled in organic–inorganic quantum wells. *Mater. Horiz.* **8**, 197–208 (2021).

5. Miller, D. A. B. *Optical Physics of Quantum Wells* (1996).
6. Milichko, V. A. *et al.* van der Waals Metal-Organic Framework as an Excitonic Material for Advanced Photonics. *Adv. Mater.* **29**, 1606034 (2017).
7. Li, Y. *et al.* Coordination assembly of 2D ordered organic metal chalcogenides with widely tunable electronic band gaps. *Nat. Commun.* **11**, 261 (2020).
8. Tran, M. *et al.* 2D coordination polymers: Design guidelines and materials perspective. *Appl. Phys. Rev.* **6**, 041311 (2019).
9. Maserati, L. *et al.* Photo-electrical properties of 2D quantum confined metal–organic chalcogenide nanocrystal films. *Nanoscale* **13**, 233–241 (2021).
10. Peach, M. E. Preparation and thermal decomposition of some metal thiophenolates. *J. Inorg. Nucl. Chem.* **41**, 1390–1392 (1979).
11. Veselska, O. & Demessence, A. d10 coinage metal organic chalcogenolates: From oligomers to coordination polymers. *Divers. Coord. Chem. Spec. Issue Honor Prof Pierre Braunstein - Part II* **355**, 240–270 (2018).
12. Smith, S. C., Bryks, W. & Tao, A. R. Supramolecular Assembly of Single-Source Metal–Chalcogenide Nanocrystal Precursors. *Langmuir* **35**, 2887–2897 (2019).
13. Huang, X., Li, J. & Fu, H. The First Covalent Organic−Inorganic Networks of Hybrid Chalcogenides:  Structures That May Lead to a New Type of Quantum Wells. *J. Am. Chem. Soc.* **122**, 8789–8790 (2000).
14. Eichhöfer, A. *et al.* Homoleptic 1-D iron selenolate complexes—synthesis, structure, magnetic and thermal behaviour of $1\infty[Fe(SeR)_2]$ (R = Ph, Mes). *Dalton Trans.* **40**, 7022–7032 (2011).
15. Eichhöfer, A. & Lebedkin, S. 1D and 3D Polymeric Manganese(II) Thiolato Complexes: Synthesis, Structure, and Properties of $\infty_3[Mn_4(SPh)_8]$ and $\infty_1[Mn(SMes)_2]$. *Inorg. Chem.* **57**, 602–608 (2018).
16. Eichhöfer, A. & Buth, G. 1-D Polymeric Iron(II) Thiolato Complexes: Synthesis, Structure, and Properties of $\infty_1[Fe(SR)_2]$ (R = Ph, Mes), $\infty_1[Fe(NH_3)(SPh)(\mu\text{-}SPh)]$ and $\infty_1[(\mu\text{-}SPh)Fe(NH_3)_2(\mu\text{-}SPh)_2Fe(\mu\text{-}SPh)]$. *Eur. J. Inorg. Chem.* **2019**, 639–646 (2019).
17. Lavenn, C. *et al.* A luminescent double helical gold(i)–thiophenolate coordination polymer obtained by hydrothermal synthesis or by thermal solid-state amorphous-to-crystalline isomerization. *J. Mater. Chem. C* **3**, 4115–4125 (2015).
18. Liu, W., Lustig, W. P. & Li, J. Luminescent inorganic-organic hybrid semiconductor materials for energy-saving lighting applications. *EnergyChem* **1**, 100008 (2019).
19. Cuthbert, H. L., Wallbank, A. I., Taylor, N. J. & Corrigan, J. F. Synthesis and Structural Characterization of $[Cu_{20}Se_4(\mu_3\text{-}SePh)_{12}(PPh_3)_6]$ and $[Ag(SePh)]_\infty$. *Z. Für Anorg. Allg. Chem.* **628**, 2483–2488 (2002).
20. Yao, K. *et al.* Strongly Quantum-Confined Blue-Emitting Excitons in Chemically Configurable Multiquantum Wells. *ACS Nano* (2020) doi:10.1021/acsnano.0c08096.
21. Yang, Z. *et al.* Ultrafast self-trapping of photoexcited carriers sets the upper limit on antimony trisulfide photovoltaic devices. *Nat. Commun.* **10**, 4540 (2019).
22. Grinblat, G. *et al.* Ultrafast All-Optical Modulation in 2D Hybrid Perovskites. *ACS Nano* **13**, 9504–9510 (2019).
23. Li, S., Luo, J., Liu, J. & Tang, J. Self-Trapped Excitons in All-Inorganic Halide Perovskites: Fundamentals, Status, and Potential Applications. *J. Phys. Chem. Lett.* **10**, 1999–2007 (2019).
24. Maserati, L., Pecorario, S., Prato, M. & Caironi, M. Understanding the Synthetic Pathway to Large-Area, High-Quality $[AgSePh]_\infty$ Nanocrystal Films. *J. Phys. Chem. C* **124**, 22845–22852 (2020).
25. Cooper, J. K. *et al.* Physical Origins of the Transient Absorption Spectra and Dynamics in Thin-Film Semiconductors: The Case of BiVO4. *J. Phys. Chem. C* **122**, 20642–20652 (2018).
26. Smith, M. D. & Karunadasa, H. I. White-Light Emission from Layered Halide Perovskites.

27. Li, J., Wang, H. & Li, D. Self-trapped excitons in two-dimensional perovskites. *Front. Optoelectron.* **13**, 225–234 (2020).
28. Williams, R. T. & Song, K. S. The self-trapped exciton. *J. Phys. Chem. Solids* **51**, 679–716 (1990).
29. Itoh, N., Stoneham, A. M. & Harker, A. H. The initial protection of defects in alkali halides: F and H centres production by non-radiative decay of the self-trapped exciton. *J. Phys. C Solid State Phys.* **10**, 4197–4209 (1977).
30. Richter, S. *et al.* The role of self-trapped excitons and defects in the formation of nanogratings in fused silica. *Opt. Lett.* **37**, 482–484 (2012).
31. Pogna, E. A. A. *et al.* Photo-Induced Bandgap Renormalization Governs the Ultrafast Response of Single-Layer MoS$_2$. *ACS Nano* **10**, 1182–1188 (2016).
32. Trovatello, C. *et al.* The ultrafast onset of exciton formation in 2D semiconductors. *Nat. Commun.* **11**, 5277 (2020).
33. Cunningham, P. D., Hanbicki, A. T., McCreary, K. M. & Jonker, B. T. Photoinduced Bandgap Renormalization and Exciton Binding Energy Reduction in WS$_2$. *ACS Nano* **11**, 12601–12608 (2017).
34. Chen, Z. *et al.* Band Gap Renormalization, Carrier Multiplication, and Stark Broadening in Photoexcited Black Phosphorus. *Nano Lett.* **19**, 488–493 (2019).
35. Price, M. B. *et al.* Hot-carrier cooling and photoinduced refractive index changes in organic–inorganic lead halide perovskites. *Nat. Commun.* **6**, 8420 (2015).
36. Yao, K. *et al.* Optically Discriminating Carrier-Induced Quasiparticle Band Gap and Exciton Energy Renormalization in Monolayer MoS$_2$. *Phys. Rev. Lett.* **119**, 087401 (2017).
37. Landsberg, P. T. *Recombination in Semiconductors.* (Cambridge University Press, 1992). doi:10.1017/CBO9780511470769.
38. Jiang, Y. *et al.* Reducing the impact of Auger recombination in quasi-2D perovskite light-emitting diodes. *Nat. Commun.* **12**, 336 (2021).
39. Seo, M. *et al.* Ultrafast Optical Microscopy of Single Monolayer Molybdenum Disulfide Flakes. *Sci. Rep.* **6**, 21601 (2016).
40. Senty, T. R., Cushing, S. K., Wang, C., Matranga, C. & Bristow, A. D. Inverting Transient Absorption Data to Determine Transfer Rates in Quantum Dot–TiO2 Heterostructures. *J. Phys. Chem. C* **119**, 6337–6343 (2015).
41. Chernikov, A., Ruppert, C., Hill, H. M., Rigosi, A. F. & Heinz, T. F. Population inversion and giant bandgap renormalization in atomically thin WS2 layers. *Nat. Photonics* **9**, 466–470 (2015).
42. Dendzik, M. *et al.* Observation of an Excitonic Mott Transition through Ultrafast Core-cum-Conduction Photoemission Spectroscopy. *Phys. Rev. Lett.* **125**, 096401 (2020).
43. Steinhoff, A. *et al.* Exciton fission in monolayer transition metal dichalcogenide semiconductors. *Nat. Commun.* **8**, 1166 (2017).
44. Ceballos, F., Cui, Q., Bellus, M. Z. & Zhao, H. Exciton formation in monolayer transition metal dichalcogenides. *Nanoscale* **8**, 11681–11688 (2016).
45. Guo, Z., Wu, X., Zhu, T., Zhu, X. & Huang, L. Electron–Phonon Scattering in Atomically Thin 2D Perovskites. *ACS Nano* **10**, 9992–9998 (2016).
46. Paritmongkol, W., Powers, E. R., Dahod, N. S. & Tisdale, W. A. Two Origins of Broadband Emission in Multilayered 2D Lead Iodide Perovskites. *J. Phys. Chem. Lett.* **11**, 8565–8572 (2020).

# Exciton self-trapping causes picoseconds recombination in metal-organic chalcogenides quantum wells

Christoph Kastl[†§], Adam M. Schwartzberg[†], and Lorenzo Maserati[†‡]

† The Molecular Foundry, Lawrence Berkeley National Laboratory, Berkeley, California 94720, USA.

§ Walter Schottky Institute and Physik Department, Technical University of Munich, Garching 85748, Germany.

‡ Center for Nano Science and Technology @PoliMi, Istituto Italiano di Tecnologia, 20133 Milan, Italy.

Correspondence to: christoph.kastl@wsi.tum.de, amschwartzberg@lbl.gov, lmaserati@lbl.gov.

**Table of contents**

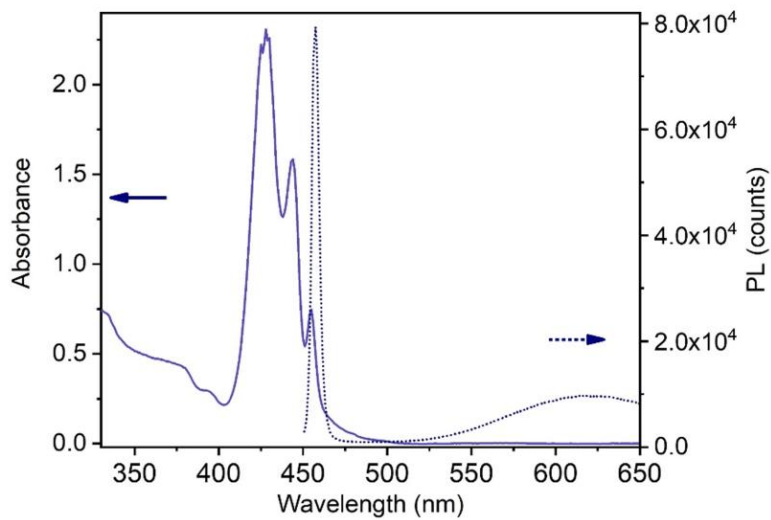

**Figure S1.** Absorbance and photo-luminescence (PL) spectra of [AgSePh]$_\infty$ film at 77 K. The absorption at 429 nm is about 1.22 times stronger than absorption at 330 nm. The PL comprises a narrow blue emission, corresponding to $X_1$ (with a Stokes shift of 14 meV), and a broad feature, indicative of trapped state emission, past 600 nm. The latter is not present in the steady state absorption spectrum.

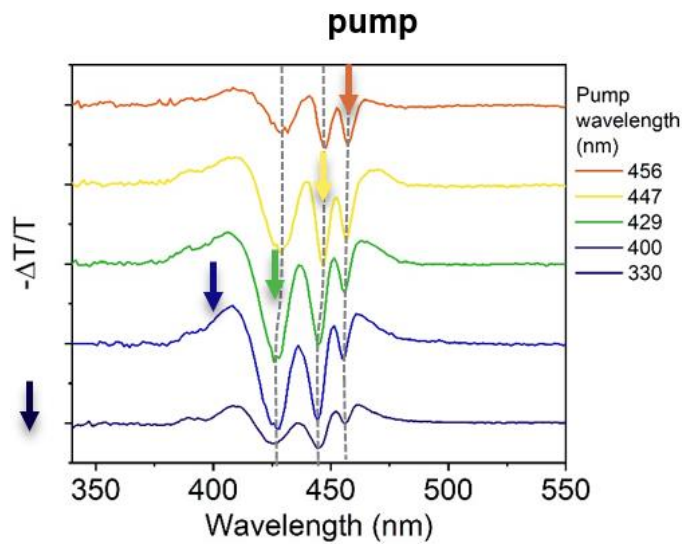

**Figure S2.** The overall shape of the transient absorption spectra directly after the pump pulse ($t \sim 200$ fs) depends only weakly on pump wavelength (indicated by the arrows). In particular, even for pumping only the lowest energy transition ($X_1$, pump wavelength 456 nm), a collective response of all excitons is observed. The latter can be understood by renormalization effects. $T_{bath} = 77$ K.

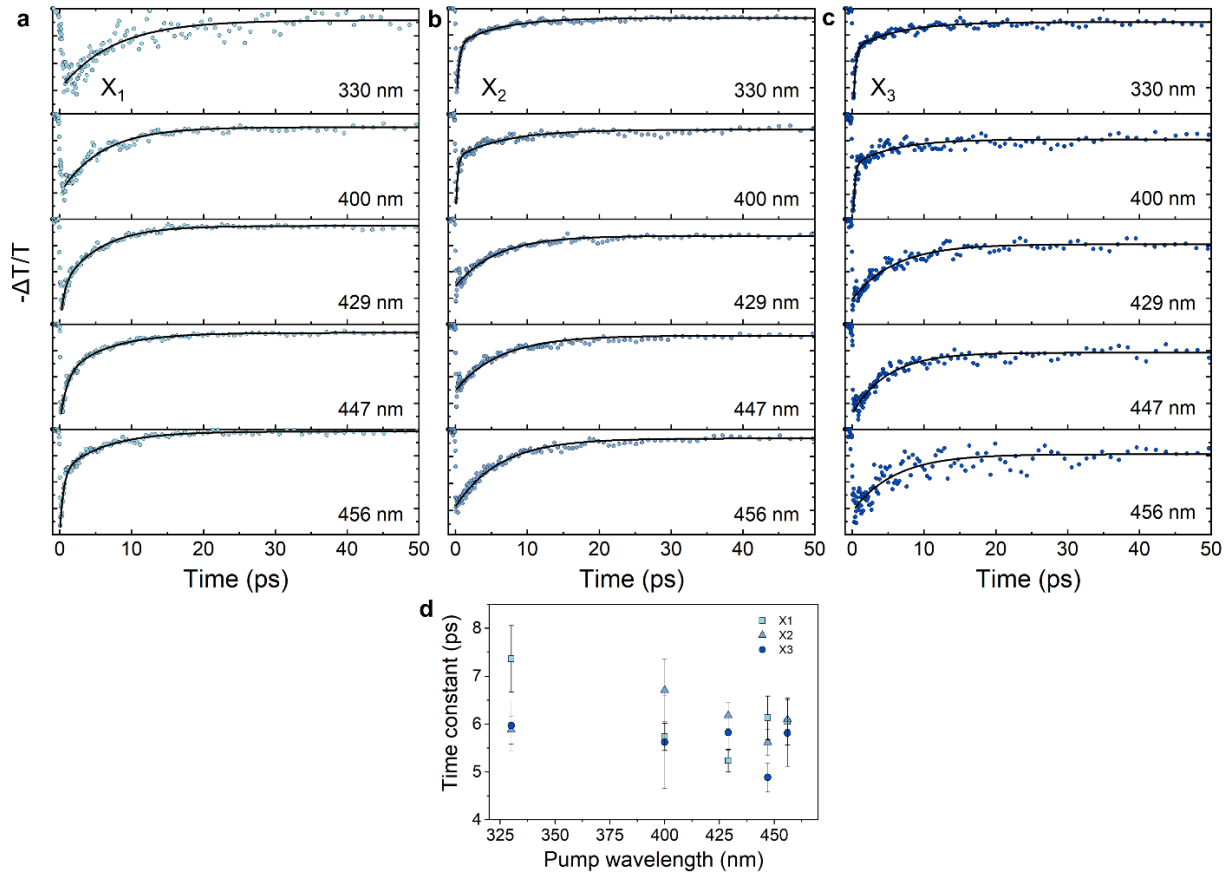

**Figure S3.** Exciton dynamics at different pump wavelengths. Dynamics of -ΔT/T at fixed wavelengths corresponding to the three excitonic transitions. Data are depicted by dots and exponential fits are depicted by solid lines: **a**, $X_1$, 427 nm. **b,** $X_2$, 447 nm. **c,** $X_3$, 456 nm. All experiments were performed at 77 K and 0.02 µJ pulse energy. d) Summary of exponential decay time constants at different pump wavelengths. The fast (sub-ps) components, which are related to hot carrier thermalization, are omitted.

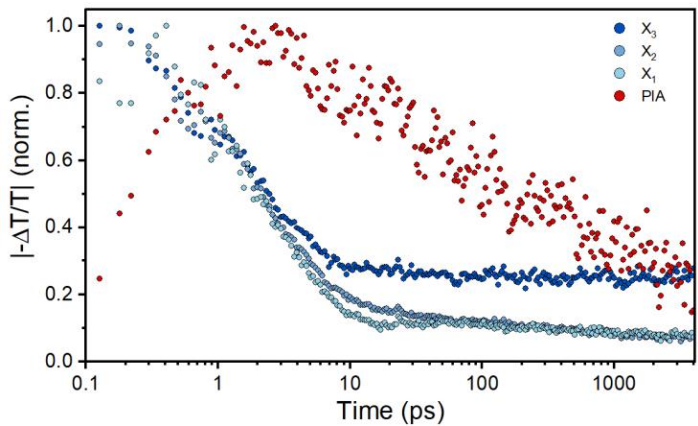

**Figure S4.** Dynamics extracted at the position of the excitonic transitions $X_1$, $X_2$, $X_3$ as well as at the photoinduced absorption band for direct excitonic pumping (pump wavelength 330 nm, pump energy 0.2 µJ, $T_{bath}$ = 77 K).

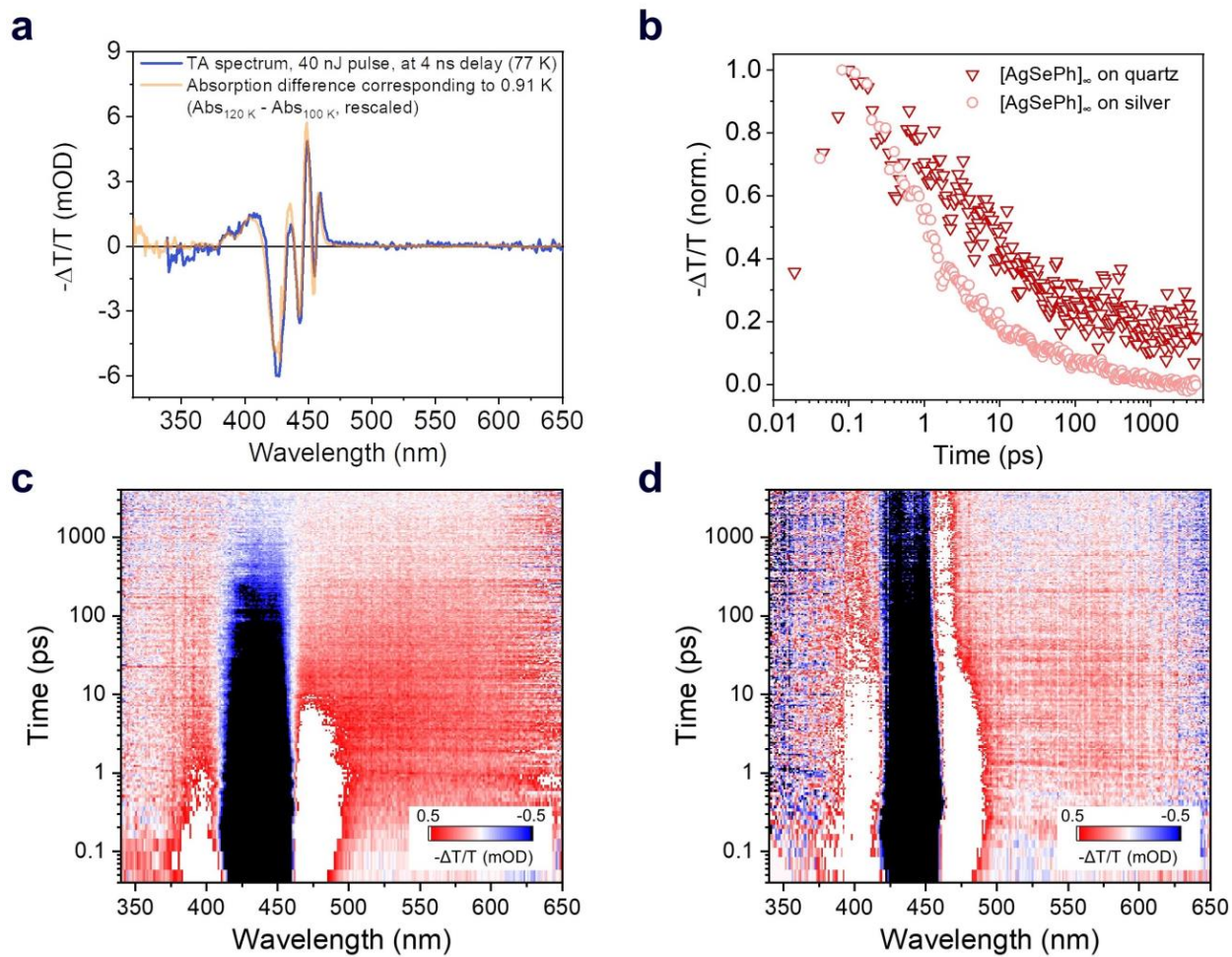

**Figure S5.** Heated phonon bath and cooling to substrate. **a,** At long time scales (~ns), the transient absorption spectrum (blue) is very well described by temperature induced changes of the steady state absorption spectrum (orange) **b,** Transient absorption dynamics extracted at the position of the excitonic transitions $X_3$, for thermally insulating substrate, i.e. quartz (dark red triangles) and thermally conducting substrate, i.e. silver film (light red circles). The thermal response persists longer (~ns) on a thermally insulating substrate compared to a thermally conductive one (~ 100 ps). **c,d** Room temperature transient absorption spectra of [AgSePh]$_\infty$ films, on silver **(c)** and quartz **(d)** substrates, respectively, pump wavelength 430 nm, pump energy 0.02 µJ.

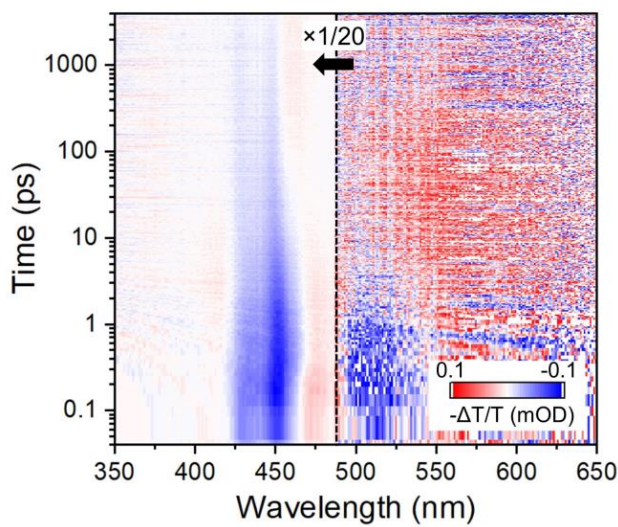

**Figure S6.** Room temperature transient absorption of [AgSePh]$_\infty$ single crystal. The single crystal shows a photoinduced absorption band around 550 nm as well. Pump wavelength 430 nm, pump energy 0.004 µJ pump energy (about 8.5 µJ/cm$^2$ due to different probe spot size compared to all other TA measurements on extended [AgSePh]$_\infty$ films).

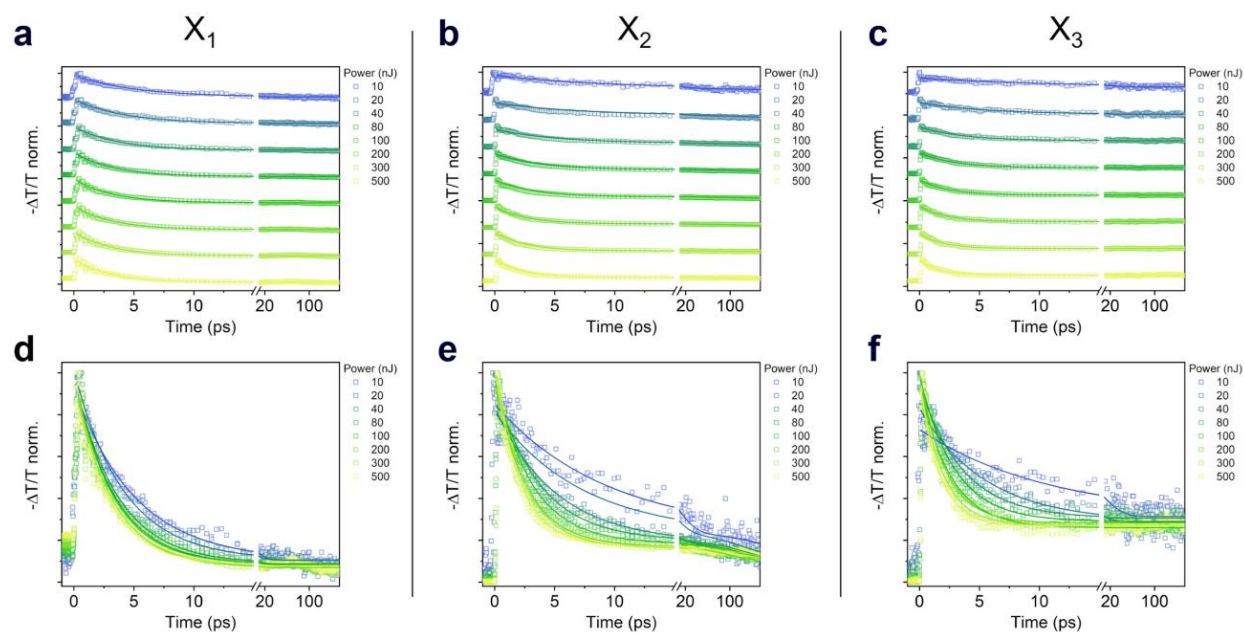

**Figure S7.** Transient absorption dynamics obtained by pumping at 429 nm, restricted to the negative signals at selected wavelengths, each representing the evolution of excitonic transitions. **a,** and **d,** $X_1$, 456 nm. **b,** and **e,** $X_2$, 447 nm. **c,** and **f,** $X_3$, 427 nm. $T_{bath}$ = 77 K.

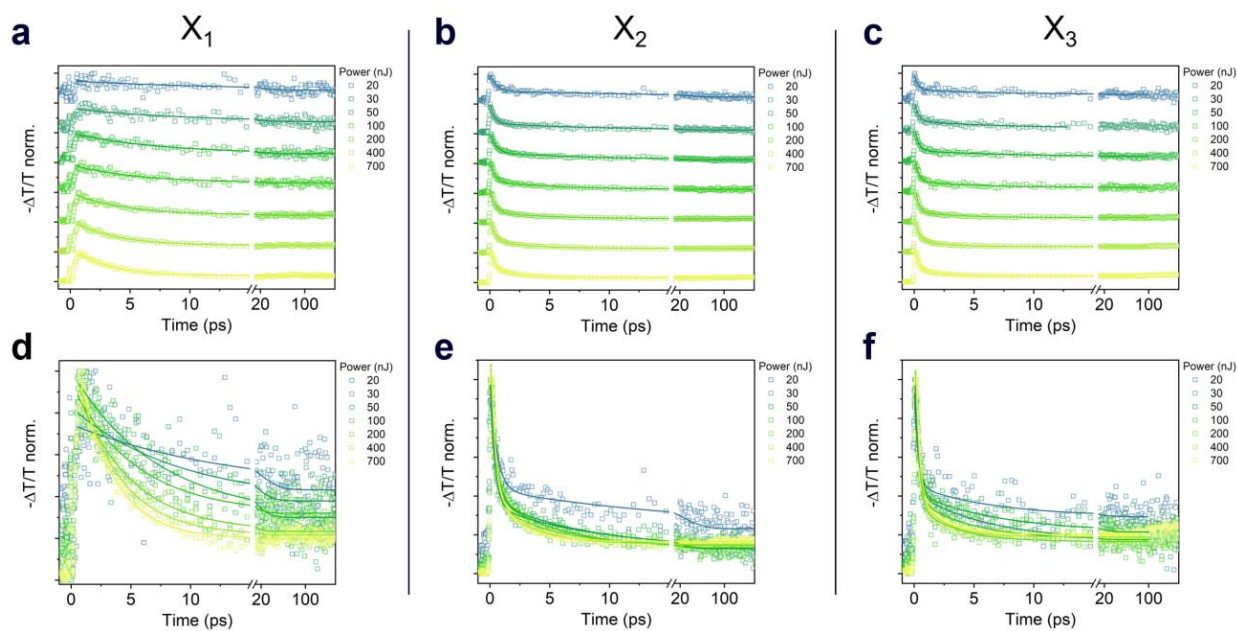

**Figure S8.** Transient absorption dynamics obtained by pumping at 330 nm, restricted to the negative signals at selected wavelengths, each representing the evolution of excitonic transitions. **a,** and **d,** $X_1$, 456nm. **b,** and **e,** $X_2$, 447 nm. **c,** and **f,** $X_3$, 427 nm. $T_{bath}$ = 77 K.

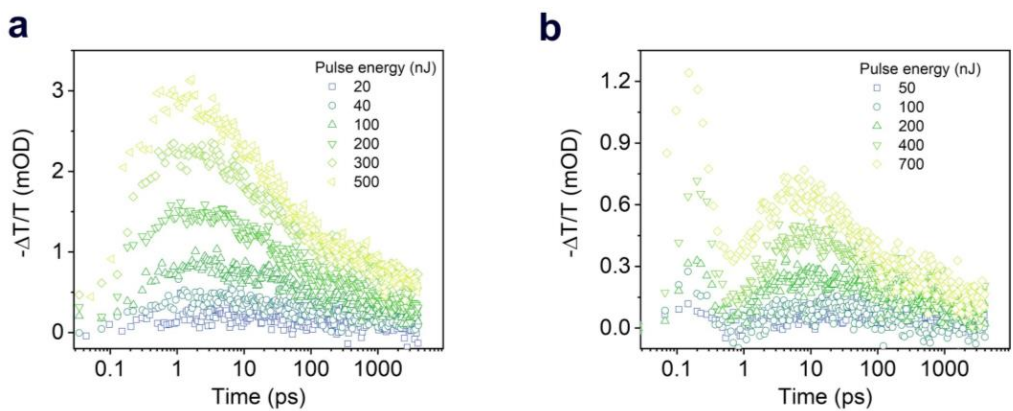

**Figure S9**. PIA transient absorption signals at 550 nm at different pump fluences. Pump wavelength **a,** 429 nm and **b,** 330 nm. $T_{bath}$ = 77 K.
19

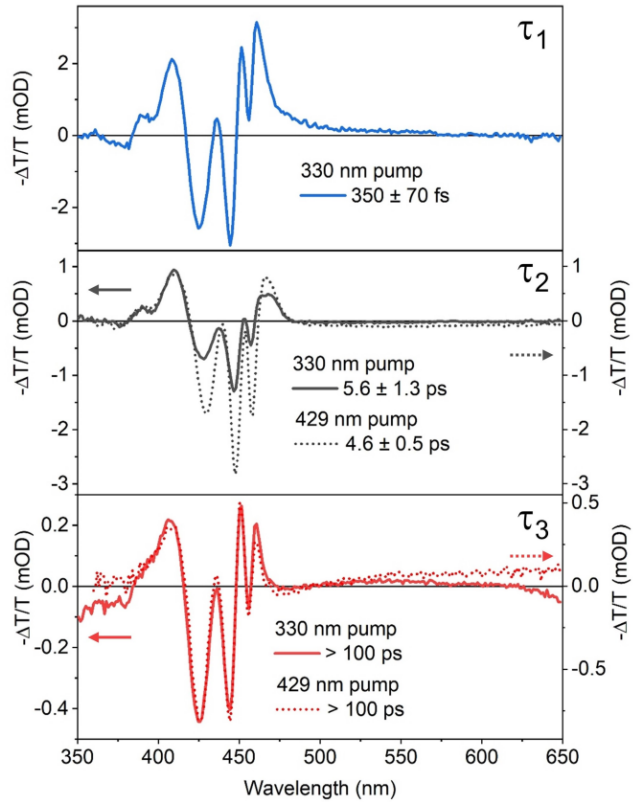

**Figure S10**. Global fit analysis of transient absorption dynamics. The different panels show the different exponential time scales ($\tau_1$, $\tau_2$, $\tau_3$) and the corresponding spectra extracted from a global fit analysis for free carrier pumping (pump wavelength 330 nm, solid lines) and resonant excitonic pumping (429 nm, dashed lines). For free carrier pumping, an initial ultrafast relaxation takes place (top panel, $\tau_1$ = 350 fs), which is absent for excitonic pumping. The intermediate time scale ($\tau_2 \sim$ 5 ps) is governed by exciton trapping, with similar spectral signatures, but increased bleaching for excitonic pumping (middle panel). On long time scales ($\tau_3 >$ 100 ps), the response is governed by a heated phonon bath, which is the same for free carrier and excitonic pumping. Pump wavelengths 330 nm and 429 nm have pump energies 0.1 µJ and 0.04 µJ, respectively. $T_{bath}$ = 77 K.

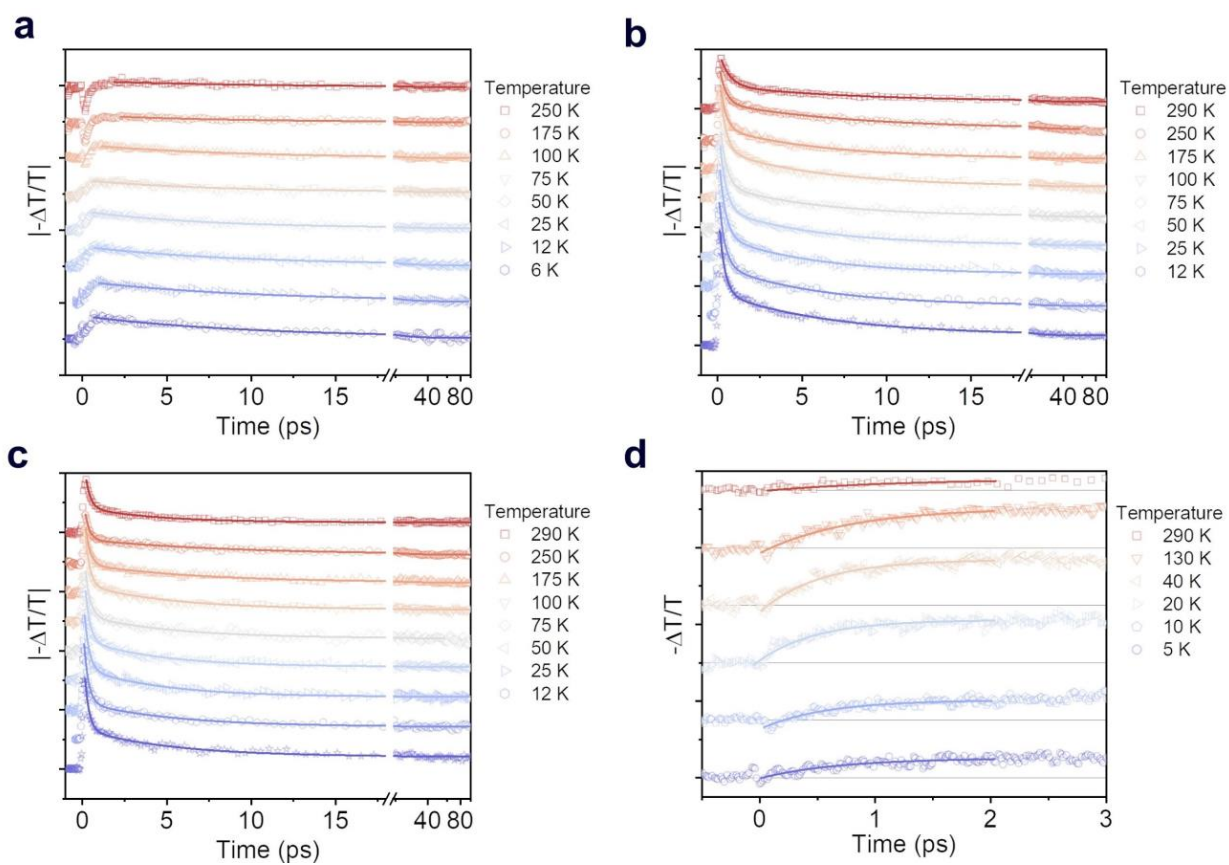

**Figure S11**. Exciton dynamics over temperature. **a-c,** Free excitons decay signals for $X_1$ **(a)**, $X_2$ **(b)**, $X_3$ **(c)**, pump wavelength 330 nm, pump energy 0.02 µJ. **d**, Self-trapped exciton rise (evaluated at 550 nm), pump wavelength 425 nm, pump energy 0.1 µJ.